\documentclass[iop]{emulateapj}
\usepackage{soul}
\usepackage{amsmath,amsfonts,amssymb,mathrsfs,graphicx}
\usepackage[squaren]{SIunits}
\usepackage{natbib}
\usepackage{multirow}
\usepackage{hyperref}
\usepackage{xcolor}
\hypersetup{
	colorlinks,
	linkcolor={red!50!black},
	citecolor={blue!50!black},
	urlcolor={blue!80!black}
}

\newcommand{\spitzer}{{\small \it Spitzer}}
\newcommand{\iso}{{\small \it ISO}}
\newcommand{\iue}{{\small IUE}}

\newcommand{\hst}{{\small {\it HST}}}

\newcommand{\Teff}{$T_{\text{eff}}$}

\begin{document}
\title{Observational Evidence Linking Interstellar UV Absorption to PAH Molecules}

\author{Avi Blasberger}

\affiliation{Department of Physics, Technion, Israel}

\author{Ehud Behar}
\affiliation{Department of Physics, Technion, Israel}
\affiliation{Department of Astronomy, University of Maryland, College Park }

\author{Hagai B. Perets}
\affiliation{Department of Physics, Technion, Israel}

\author{Noah Brosch}
\affiliation{The Wise Observatory and School of Physics and Astronomy, Tel Aviv University, Israel }

\author{Alexander G.G.M. Tielens}
\affiliation{Leiden Observatory, Leiden University, The Netherlands}

\begin{abstract}
	
The 2175\,\AA\ UV extinction feature was discovered in the mid-1960s, yet its physical origin remains poorly understood.
One suggestion is absorption by Polycyclic Aromatic Hydrocarbons (PAH) molecules, which is supported by theoretical molecular structure computations and by laboratory experiments. 
PAHs are positively detected by their 3.3, 6.2, 7.7 8.6, 11.3 \& 12.7\,$\mu$m IR emission bands, which are specified by their modes of vibration. 
A definitive empirical link between the 2175\,\AA\ UV extinction, and the PAH IR emission bands, however, is still missing.
We present a new sample of hot stars that have both 2175\,\AA\ absorption and PAH IR emission.
We find significant shifts of the central wavelength of the UV absorption feature, up to 2350\,\AA , but predominantly in stars that also have IR PAH emission.
These UV shifts depend on stellar temperature in a fashion that is similar to the shifts of the 6.2 and 7.7$\mu$m PAH IR bands, namely the features are increasingly more red-shifted as the stellar temperature decreases, but only below $\sim 15$\,kK.  
Above 15\,kK both UV and IR features retain their nominal values.
Moreover, we find a suggestive correlation between the UV and IR shifts.
We hypothesize that these similar dependences of both the UV and IR features on stellar temperature hint to a common origin of the two in PAH molecules and may establish the missing link between the UV and IR observations. 
We further suggest that the shifts depend on molecular size, and that the critical temperature of $\sim 15$\,kK above which no shifts are observed is related to the onset of UV driven hot-star winds and their associated shocks.

\end{abstract}

\maketitle

\flushbottom


\section{Introduction}
The 2175\,\AA\ UV extinction feature was discovered in 1965 by \citet{Stecher 1} with sounding rocket observations.  
Since then, several satellite observatories helped to characterize it towards many sightlines
\citep{Savage85,FM86,FM88,FM90,FM05,FM07,CCM89,valancic2004}, yet its physical origin is still being debated.
One suggested origin is absorption by Polycyclic Aromatic Hydrocarbons (PAH) molecules, that are the most abundant organic molecules in nature and a key ingredient of the interstellar medium (ISM)  \citep{Tielens1}. 
Indeed, theoretical molecular structure computations \citep{leger 1,Puget 1,Draine  1,zubko  1} as well as by laboratory experiments \citep{Steglich 1, Steglich 2} indicate that PAHs absorb ultraviolet (UV) light and emit in broad Mid IR (MIR) bands. These works and other  have ascribed the 2175\,\AA\ feature to very small graphite grains \citep{leger 1,Puget 1,Draine  1,zubko  1,Stecher 2}. 

PAHs are organic molecules composed of C and H atoms, structured in the form of multiple aromatic rings with delocalized electrons.
They contain up to 10\% of the C content in our galaxy, and have thus been hypothesized to play a key role in the earliest forms of organic life in the universe \citep{Ehrenfreund 1}. 
The PAHs are ubiquitous in the ISM, as revealed by their emission bands in the MIR around 3.3, 6.2, 7.7, 8.6, 11.2, and 12.7 $\mu$m, originating in different vibrational modes \citep{Allamandola89}. 
The typical temperature of ISM dust grains ($<$ 100 K) cannot account for efficient MIR emission, which is thus attributed to smaller species with low heat capacity, such as PAHs. 
The excitation energies of $\sim 10$\,eV then imply typical molecule sizes of about 50 C-atoms \citep{Sellgren 1}	
The PAH hypothesis is, thus, that the MIR PAH emission is excited by UV photons in the ISM, although perhaps not exclusively \citep{lidr2002, smith2004}. 

Mid IR observations with the \iso\ and \spitzer\ space telescopes revealed the PAH emission band profiles, and their relative intensities \citep{van Diedenhoven,Hony 2006,Galliano,mori1}. 
Observations showed a dependence of the central wavelength of the 6.2, 7.7, and 8.6 $\mu$m bands on the illuminating stellar temperature \citep{Sloan 1,Acke 1}.
The variation in band position has been attributed to varying size and charge of the PAH molecules \citep{pino, Ricca 1}. 
The position of the UV extinction feature 2175\,\AA , on the other hand, is rather uniform.
This contrasts with the diversity of PAHs \citep{joblin 1}, ISM conditions, and thus challenges the PAH hypothesis.
Nonetheless, a small number of cases have been reported where the UV extinction feature occurs at longer wavelengths \citep{Greenstein  1,Hecht  1,Buss 1}.

Here, we use an analysis of a new sample of stars that have both UV and MIR spectra to demonstrate the missing connection between the UV absorption and MIR emission bands. 
The remainder of this paper is organized as follows: In Sec.~\ref{sample} we describe the sample, in Sec.~\ref{method} we detail the measurements, in Sec.~\ref{results} we present the results, and in Sec.~\ref{discussion} we conclude with a discussion and  draw the conclusion in Sec.~\ref*{Conclusions}.

\section{The sample}
\label{sample}
We assembled a sample of objects of spectral type A2 and earlier (hotter) that have spectra both from the \iue\ (UV), and from either the \spitzer\ or \iso\ (MIR) telescopes.
We did not impose any visible extinction requirements; therefore, the present sample includes many stars with low color excess that would not be included in studies aimed at obtaining general extinction curves.
We identified 27 stars that have both significant (Equivalent Width $\rm{EW} \ge 100$\,\AA) UV absorption around 2175\,\AA, and significant PAH emission ($\rm{EW} \ge 0.1\mu$m in the 7.7 $\mu$m band).
Although not selected for it, most stars in the present sample happen to be pre main-sequence stars that are shrouded in dust and gas, often in the form of a proto-planetary disk, which has been shown to provide favorable conditions for the growth of large PAHs \citep{Berné 1}. 
Besides its main attribute of including both UV and IR spectra, the present sample is unique in that it extends both to cooler stars (down to 8.25 kK) than those usually studied in the UV, and to hotter stars (up to 37 kK) than those usually studied in the IR. 

In order to test the UV-MIR connection and its relation to PAHs, we also constructed a control sample of 32 stars with UV absorption ($\rm{EW} \ge 100$\,\AA), but with no detectable MIR emission ($\rm{EW} \le 0.1\,\mu$m at 7.7 $\mu$m), for comparison.
A list of the present sample can be found in Table~1; objects 1-27 are those with both UV absorption and PAH MIR emission, while objects 28-59 are those with UV absorption but no detectable MIR emission.
Spectral type and B \& V magnitudes were taken from the SIMBAD astronomical database\footnote{http://simbad.u-strasbg.fr/simbad/}.
Stellar temperatures \Teff\ were determined from  \citet{Pecaut} according to spectral type.
Given the poor knowledge of the bolometric corrections, $T_{\text{eff}}$ is uncertain by a few percent \citep{dejager}, which is insignificant here.
Intrinsic B$_0$ -V$_0$ values for main sequence stars were taken from the Space Telescope Science Institute (STScI) 2002 archive \footnote{http://www.stsci.edu/~inr/intrins.html}, For other spectral types we used \citet{Schmidt-Kaler} tables.

Despite the fact that the  Hubble Space Telescope (\hst ) has improved UV capabilities compared to \iue , had  it observed only about 10\% as many UV-bright stars as IUE.
For example, only two stars from our sample have adequate HST spectra therefore we used the IUE data. Moreover,  the two cases in which the HST and IUE had the same objects the analyzed results was similar. 

\section{Data reduction and analysis}
\label{method}

\subsection{UV}

The \iue\ spectra were extracted from the Mikulski Archive for Space Telescopes (MAST)\footnote{https://archive.stsci.edu/iue/} , and comprise of data obtained by three cameras. The short wavelength camera (SWP) includes spectra from 1150\,\AA\ to 1978\,\AA, the two long wavelength cameras (LWP \& LWR) include spectra from 1850\,\AA\ to 3347\,\AA, in both the large aperture and the small aperture. Observations that include at least one SWP spectrum and one LWP or LWR spectrum were selected. 
In the combined spectrum, the less noisy SWP is used up to 1970\,\AA, and LWP/LWR is used above 1970\,\AA.
Where possible, we used the large \iue\ aperture since the small aperture can suffer from large flux variations during the observation. 
\iue\ bins flagged for bad quality were ignored. 
Observations that include more than 20\,\AA\ of sequentially flagged data, or more than 60\,\AA\ in total in the wavelength range of interest were omitted from the sample.
In cases where several observations exist, we combined spectra bin-by-bin to obtain error-weighted (1/$\sigma^2$) means and uncertainty of the mean (1/$\sigma^2$ added in quadrature). In cases where the large aperture data did not comply with the above set of quality requirements, and small aperture spectra exist, we completed the large aperture spectrum with the small aperture data, using a flux scaling based on the overlap regions.  

\begin{figure}[]
	\hspace{-1cm}
	\includegraphics[width=0.6\textwidth]{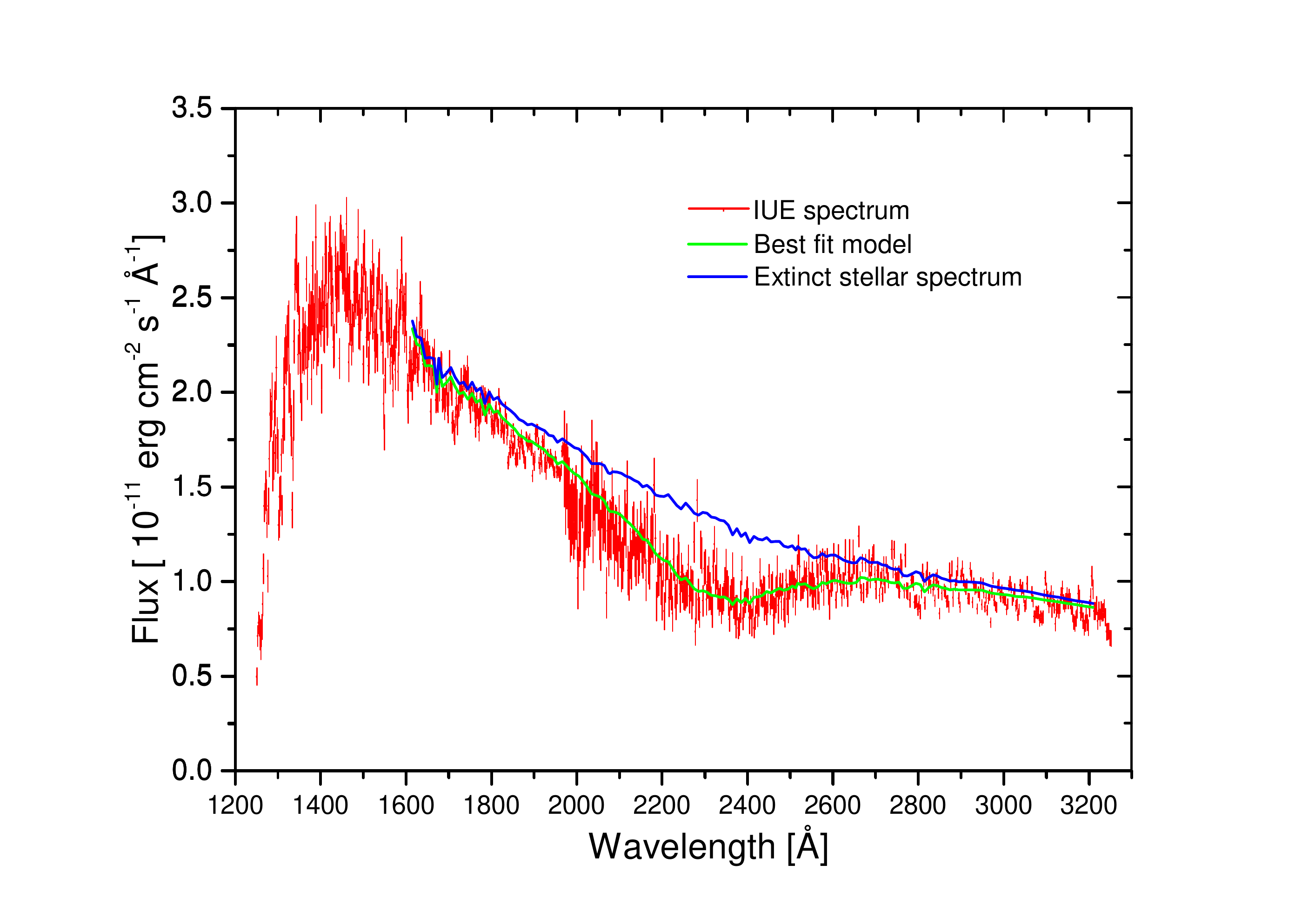}
	\caption{IUE spectrum and fitted model demonstrating significantly shifted absorption. 
		Red data points are the IUE spectrum of HD 100546, the blue line is the stellar continuum from the model of \citep{Castelli 1} modified for extinction. The green line denotes the best-fitted model that includes also the Drude-profile of the UV feature, whose central wavelength in HD 100546 is 2301\AA }
	\label{UVfit}
\end{figure}

The combined spectra were fitted between 1600\,\AA\ and 3200\,\AA.  
We fitted the \iue\ spectra directly and locally around the 2175\,\AA\ feature using the \citet{Castelli 1} stellar-atmospheric spectra, modulated by a power law to represent the broadband extinction, and absorbed by a feature with a Drude profile. 
The measured flux spectrum $F_\lambda$ as a function of wavelength $\lambda$ then takes the following form:

\begin{equation} 
\label{absorption}
F_\lambda=F_{CK}N \left( \frac{\lambda}{2000 {\rm \AA } } \right)^{-\alpha} \exp \left( \frac{-A}{\pi}\frac{\lambda^2}{(\lambda^2-\lambda_0^2) +\lambda^2w^2} \right)
\end{equation}

\noindent where $F_{CK}(\lambda)$ is the theoretical stellar spectrum \citep{Castelli 1}, 
$N$ is the normalization, and the power law slope is $-\alpha$.
In the exponent, $A$ represents the amplitude of the absorption feature, $w$ its width, and $\lambda_0$ its central wavelength.
The optical depth at the central wavelength is $\tau(\lambda_0)$ = A/($\pi$w$^2$). 
Eq. \ref{absorption} is fitted to the measured spectrum of each object by minimizing $\chi^2$.
The best-fit values are consequentially obtained for the parameters $N, \alpha,w$, and $\lambda_0$. 
For the stellar spectra, we used models of \citet{Castelli 1}, based on spectral type and the metallicity that best fits the \iue\ spectrum.
Most stars are fitted extremely well, while the few (3/62) that are not (reduced $\chi^2$ /d.o.f $\ge$ 5) were eliminated from the sample.  
In Figure \ref{UVfit}, we present a typical fit. 

We compute the equivalent width ($\rm{EW}$),by integrating as follows:

\begin{equation} \label{EW}
EW=  \frac{\int_{1600\AA}^{3200\AA}(F_\lambda^0-F_\lambda)d\lambda}{F_\lambda^0(\lambda_0)}
\end{equation}

\noindent where $F_\lambda^0$ is the unabsorbed continuum (Eq. \ref{absorption} with $A = 0$).
For the present study of PAH UV absorption, we retained only objects with 
$\rm{EW} \ge 100$\,\AA. The results are listed in Table \ref*{all stars}. 

\begin{table*}
	\caption{List of stars, spectral properties, and UV fitted parameters}
	\label{all stars}
	\centering
	\begin{tabular}{c c c c c c c }
		\hline\hline		
Object & Object  & Spectral & E(B-V)~$^b$  & $\lambda_0\pm \Delta \lambda_0$[\AA] & $\chi^2$/d.o.f.  & EW      \\
		\# & Name       & Type~$^a$       &              &2$\sigma$ errors & & (\AA ) \\
		\hline
		1  & RR-Tau     & A0:IVe      & 0.83  & 2307.2$\pm$24       & 2.157  & 867.5   \\
		2  & HD95881    & A1/A2III/IV & 0.13  & 2400.2$\pm$12.8     & 4.734  & 352     \\
		3  & HD141569   & B9.5V       & 0.09  & 2209.2$\pm$7.6      & 2.038  & 169.9   \\
		4  & HD31293    & A0Ve        & 0.13  & 2261.1$\pm$13.9     & 4.338  & 150.7   \\
		5  & HD179218   & A0Ve        & 0.11  & 2247.8$\pm$7.5      & 3.677  & 232.9   \\
		6  & HD97048    & A0Vep       & 0.31  & 2218.1$\pm$2.9      & 0.813  & 483.4   \\
		7  & HD36917    & B9III/IV    & 0.17  & 2253.2$\pm$8.5      & 2.625  & 254.9   \\
		8  & HD89353    & B9.5Ib-II   & 0.37  & 2368.6$\pm$14       & 1.99   & 252.7   \\
		9  & HD97300    & B9V         & 0.24  & 2203.1$\pm$4.4      & 1.098  & 405.1   \\
		10 & HD100546   & B9Vne       & 0.48  & 2301.3$\pm$8.6      & 2.373  & 169.7   \\
		11 & HD139636   & B8/9III     & 0.29  & 2245.7$\pm$5.1      & 1.711  & 542.8   \\
		12 & BD+30549   & B8:p        & 0.59  & 2177  $\pm$2.9      & 1.587  & 529.9   \\
		13 & HD44179    & B9Ib/II     & 0.34  & 2364.6$\pm$15.3     & 4.395  & 672.6   \\
		14 & V699Mon    & B7IIne      & 0.7   & 2241.7$\pm$4.8      & 2.708  & 762.4   \\
		15 & HD281159   & B5V         & 0.83  & 2189.6$\pm$1.8      & 1.6    & 911.1   \\
		16 & HD124237   & B5/B6V      & 0.48  & 2179.9$\pm$2.5      & 2.522  & 643.8   \\
		17 & HD147889   & B2III/IV    & 1.05  & 2179.3$\pm$1.1      & 2.259  & 1023.7  \\
		18 & BD+404124  & B2Ve        & 0.98  & 2216.6$\pm$3        & 4.493  & 849.1   \\
		19 & HD200775   & B2Ve        & 0.61  & 2193.8$\pm$3.7      & 1.725  & 481     \\
		20 & HD37903    & B3IV        & 0.31  & 2183.4$\pm$2.7      & 1.259  & 359.2   \\
		21 & HD36982    & B1.5Vp      & 0.37  & 2201.9$\pm$3.5      & 1.364  & 402.5   \\
		22 & cd-4911554 & B1Iae       & 0.59  & 2154.7$\pm$4.6      & 2.419  & 353.3   \\
		23 & SK -66 19  & OB          & 0.43  & 2197.4$\pm$9        & 3.774  & 178.8   \\
		24 & HD37020    & B0V         & 0.32  & 2215.2$\pm$10.4     & 2.416  & 182.6   \\
		25 & cd-4211721 & B0IVe       & 1.58  & 2160.3$\pm$6.2      & 1.876  & 1149.8  \\
		26 & HD37022    & O7Vp        & 0.34  & 2215.1$\pm$4.7      & 1.051  & 190.7   \\
		27 & HD213985   & A0III       & 0.16  & 2285  $\pm$4.5      & 2.45   & 910     \\
		28 & HD105209   & A1V         & 0.18  & 2268  $\pm$15.1     & 2.225  & 212.4   \\
		29 & HD147009   & A0V         & 0.28  & 2233  $\pm$3.3      & 1.888  & 488.8   \\
		30 & HD163181   & O9.5Ia/ab   & 0.65  & 2170.6$\pm$2.3      & 3.282  & 691.5   \\
		31 & HD149914   & B9.5IV      & 0.26  & 2212  $\pm$4.2      & 1.736  & 477.6   \\
		32 & HD145554   & B9V         & 0.2   & 2224.9$\pm$4.2      & 2.5    & 342.8   \\
		33 & HD58647    & B9IV        & 0.14  & 2217.2$\pm$7.6      & 2.139  & 189.7   \\
		34 & HD147701   & B5III       & 0.71  & 2176.2$\pm$3.2      & 2.395  & 726.4   \\
		35 & HD27396    & B4IV        & 0.13  & 2197.9$\pm$5.7      & 2.132  & 285.6   \\
		36 & HD161573   & B4          & 0.17  & 2195.2$\pm$4.6      & 2.503  & 327.7   \\
		37 & HD38087    & B3II        & 0.31  & 2192.5$\pm$1.9      & 1.211  & 457.8   \\
		38 & HD175156   & B3II        & 0.33  & 2183  $\pm$3.3      & 4.383  & 408.8   \\
		39 & HD141926   & B2nne       & 0.74  & 2192.6$\pm$1.9      & 2.211  & 781.5   \\
		40 & BD+622125  & B5          & 0.73  & 2181.8$\pm$2.7      & 2.117  & 815.4   \\
		41 & Hiltner188 & B1V         & 0.54  & 2180.2$\pm$2.9      & 3.159  & 712.3   \\
		42 & HD154445   & B1V         & 0.38  & 2197.4$\pm$2.2      & 2.749  & 594.9   \\
		43 & BD+56524   & B1Vn        & 0.6   & 2181.7$\pm$2.2      & 1.771  & 630.7   \\
		43 & HD239722   & B2IV        & 0.78  & 2172.5$\pm$1.5      & 2.097  & 870.5   \\
		45 & HD23180    & B1III       & 0.31  & 2136.1$\pm$4.9      & 3.241  & 439.8   \\
		45 & HD77581    & B0.5Ia      & 0.73  & 2170.7$\pm$2.8      & 3.278  & 696     \\
		46 & HD239729   & B0V         & 0.66  & 2181.5$\pm$2.7      & 2.451  & 554     \\
		48 & CPD-592596 & B0V         & 0.7   & 2195.5$\pm$2.5      & 3.422  & 569.7   \\
		49 & HD147165   & O9.5V       & 0.43  & 2184.5$\pm$2.8      & 3.37   & 539.6   \\
		50 & BD+60594   & O9V         & 0.59  & 2182.8$\pm$2.8      & 3.253  & 642.2   \\
		51 & HD151804   & O8.5Iab(f)p & 0.36  & 2213.7$\pm$6.4      & 4.37   & 283.7   \\
		52 & HD149404   & O8.5Iab(f)p & 0.69  & 2203.6$\pm$2.4      & 4.043  & 616.4   \\
		53 & HD217086   & O7Vnn((f))z & 0.96  & 2200.7$\pm$1.7      & 2.008  & 855     \\
		54 & HD46150    & O5V((f))z   & 0.46  & 2203.3$\pm$2.7      & 2.587  & 505.8   \\
		55 & HD93162    & O2.5If*/WN6 & 0.04  & 2207.5$\pm$3.4      & 4.467  & 594.3   \\
		56 & HD29647    & B8III       & 1.02  & 2220.6$\pm$8.1      & 2.85   & 1097.5  \\
		57 & HD132947   & B9V         & 0.13  & 2216.9$\pm$8.5      & 2.251  & 224.4   \\
		58 & HD145631   & B9V         & 0.2   & 2186.4$\pm$3.7      & 4.006  & 287.7   \\
		59 & HD23441    & A0Vn        & 0.005 & 2239.2$\pm$11.6     & 3.413  & 130.6 \\
		\hline\hline		 
	\end{tabular}

\end{table*}

\subsection{MIR}

\begin{figure}[]
	\hspace{-0.5cm}
	\includegraphics[width=0.55\textwidth]{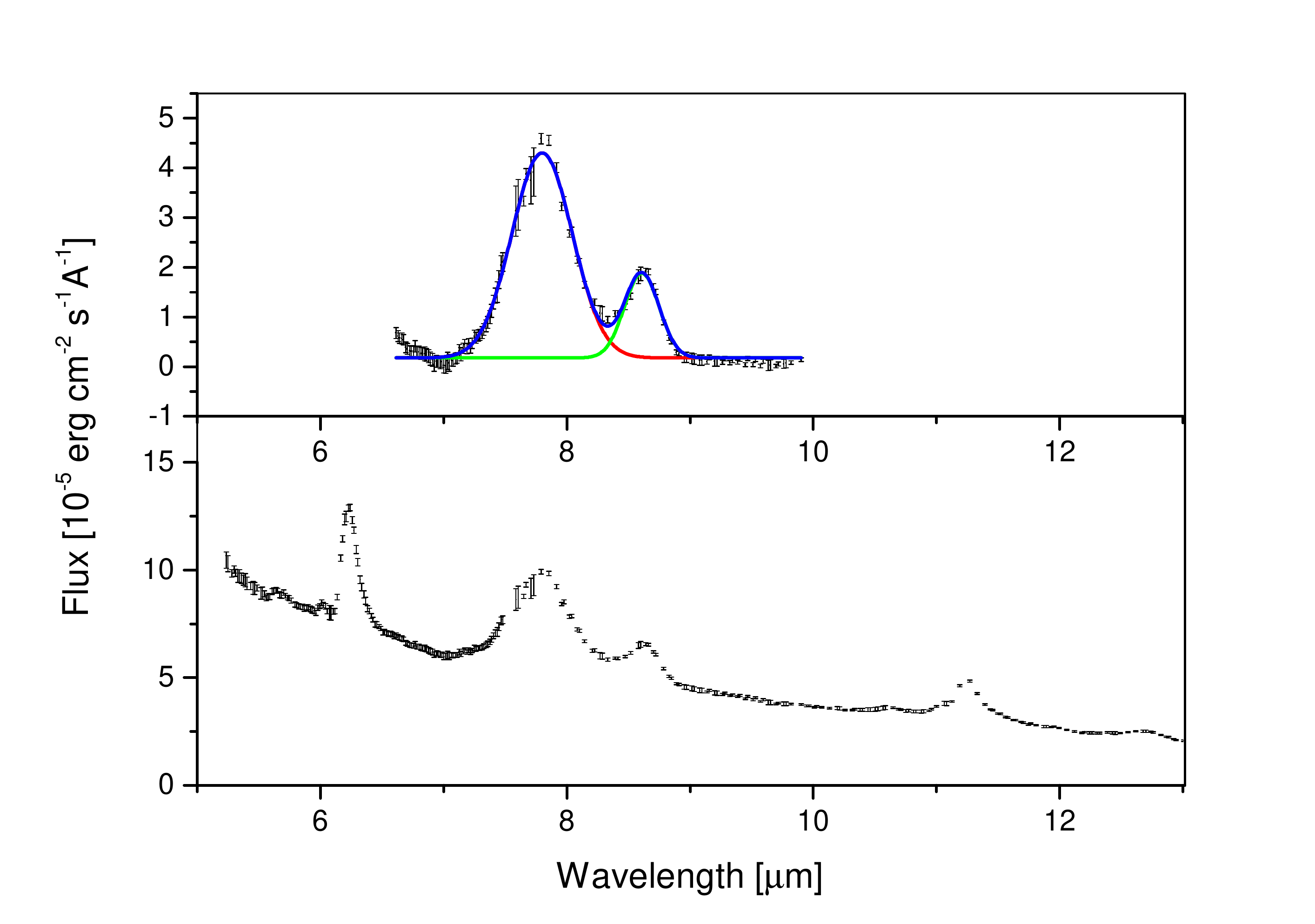}
	\caption{Spitzer spectrum and fitted model demonstrating shifted emission.
		(top panel) Two-Gaussian fit of the red-shifted 7.7 $\mu$m and 8.6 $\mu$m emission bands of RR Tau. A local linear continuum has been subtracted out.  Full Spitzer spectrum of RR Tau in which multiple PAH bands are clearly seen (bottom panel). }
	\label{IRfit}
\end{figure}

IR spectra were extracted for all the 59 UV selected stars listed in Table \ref*{all stars} from IRSA – the NASA/IPAC Infra-Red Science Archive \footnote{http://irsa.ipac.caltech.edu/frontpage/}.
We used data when available from the \spitzer /IRS instrument, and from \iso / SWS when not.
For \spitzer , we combined the spectra from different slits between 5.13$\mu$m - 7.6 $\mu$m, 7.33 $\mu$m - 8.66 $\mu$m, and 7.46 $\mu$m - 14.29 $\mu$m to obtain a continuous spectrum between 5.13 $\mu$m - 14.29 $\mu$m. For ISO, the data extracted from IRSA is already combined \citep{Sloan 2}. The brightest PAH emission bands are around 3.3, 6.2, 7.7, 8.6, and 
11.2 $\mu$m. We measured their individual fluxes by fitting Gaussians over the local continuum. We also tried Lorentzian profiles, as well as simple spline functions \citep{Galliano}. It has been shown that the resulting flux is only slightly affected by the chosen profile, while it may depend somewhat on the continuum determination \citep*{Uchida}. The central wavelength is much less affected by choice of profile function.
Since all of our data were fitted with the same method, the trends and conclusions do not change with such small systematic uncertainties.
A sample IR spectrum is shown in Figure \ref{IRfit} 

About half (27/59) of the sources in our sample have a measurable MIR emission band around 7.7 $\mu$m with an $\rm{EW} \ge$ 0.1$\,\mu$m. 
This half of the sample is referred to here as stars with MIR PAH emission,  while the other stars are referred to as being without MIR PAH emission.
For stars with MIR PAH emission, we fitted the flux and central wavelength of the 6.2\,$\mu$m and 7.7\,$\mu$m bands.
The results are listed in Table \ref*{pah stars}.

\begin{table*}[h!]
	\caption{List of stars with PAH MIR bands, and fitted parameters.  }
	\centering
		\begin{tabular}{cccccc}
		\hline\hline
		\# & Object  Name    & Telescope   & Spectral Type  & 6.2 Band center     & 7.7 Band center      \\
		&            &         &           &   $\lambda_0\pm \Delta \lambda_0$[$\mu$m]    &    $\lambda_0\pm \Delta \lambda_0$[$\mu$m]          \\
		\hline
		1  & RR-Tau     & Spitzer & A0:IVe    & 6.239$\pm$0.0021   & 7.798$\pm$0.0036   \\
		2  & HD95881    & Spitzer & A1/A2III/IV& 6.273$\pm$0.0047  & 7.932$\pm$0.009    \\
		3  & HD141569   & Spitzer & B9.5V     & 6.267$\pm$0.0069   & 7.909$\pm$0.0073   \\
		4  & HD31293    & ISO     & A0Ve      & 6.262$\pm$0.0039   & 7.789$\pm$0.0124   \\
		5  & HD179218   & ISO     & A0Ve      & 6.276$\pm$0.0011   & 7.8  $\pm$0.0012   \\
		6  & HD97048    & Spitzer & A0Vep     & 6.232$\pm$0.0017   & 7.781$\pm$0.0035   \\
		7  & HD36917    & Spitzer & B9III/IV  & 6.23 $\pm$0.0036   & 7.711$\pm$0.0081   \\
		8  & HD89353    & ISO     & B9.5Ib-II & 6.265$\pm$0.0006   & 7.823$\pm$0.0017   \\
		9  & HD97300    & Spitzer & B9V       & 6.235$\pm$0.0027   & 7.715$\pm$0.0042   \\
		10 & HD100546   & ISO     & B9Vne     & 6.25 $\pm$0.0005   & 7.901$\pm$0.0007   \\
		11 & HD139636   & Spitzer & B8/9III   & 6.257$\pm$0.0052   & 7.752$\pm$0.016    \\
		12 & BD+30549   & Spitzer & B8:p      & 6.221$\pm$0.0042   & 7.69 $\pm$0.0056   \\
		13 & HD44179    & ISO     & B9Ib/II   & 6.268$\pm$0.0005   & 7.827$\pm$0.0004   \\
		14 & V699Mon    & Spitzer & B7IIne    & 6.232$\pm$0.0035   & 7.687$\pm$0.0046   \\
		15 & HD281159   & Spitzer & B5V       & 6.219$\pm$0.0045   & 7.646$\pm$0.0173   \\
		16 & HD124237   & Spitzer & B5/B6V    & 6.225$\pm$0.0053   & 7.719$\pm$0.0109   \\
		17 & HD147889   & Spitzer & B2III/IV  & 6.221$\pm$0.0066   & 7.651$\pm$0.0089   \\
		18 & BD+404124  & Spitzer & B2Ve      & 6.221$\pm$0.0007   & 7.657$\pm$0.0017   \\
		19 & HD200775   & Spitzer & B2Ve      & 6.232$\pm$0.0035   & 7.612$\pm$0.0061   \\
		20 & HD37903    & Spitzer & B3IV      & 6.236$\pm$0.0031   & 7.659$\pm$0.007    \\
		21 & HD36982    & Spitzer & B1.5Vp    & 6.234$\pm$0.0037   & 7.659$\pm$0.0046   \\
		22 & cd-4911554 & ISO     & B1Iae     & 6.228$\pm$0.004    & 7.801$\pm$0.0027   \\
		23 & SK -66 19  & Spitzer & OB        & 6.231$\pm$0.0084   & 7.7  $\pm$0.0296   \\
		24 & HD37020    & Spitzer & B0V       & 6.229$\pm$0.0026   & 7.572$\pm$0.0241   \\
		25 & cd-4211721 & ISO     & B0IVe     & 6.227$\pm$0.0008   & 7.667$\pm$0.0012   \\
		26 & HD37022    & Spitzer & O7Vp      & 6.229$\pm$0.0027   & 7.64 $\pm$0.0055   \\
		27 & HD213985   & Spitzer & A0III     & -----              & 7.83 $\pm$0.0017   \\
		\hline\hline
	\end{tabular}
	\label{pah stars}
\end{table*}

\section{results}
\label{results}
We find no particular pattern or correlation with feature strength ($\rm{EW}$) either in the UV or in the MIR. We thus focus on the central wavelength $\lambda_0$ (i.e., position) of the features, which we show below to be highly revealing.

\begin{figure}[!]
	\hspace{-1cm}
	\includegraphics[width=0.59\textwidth]{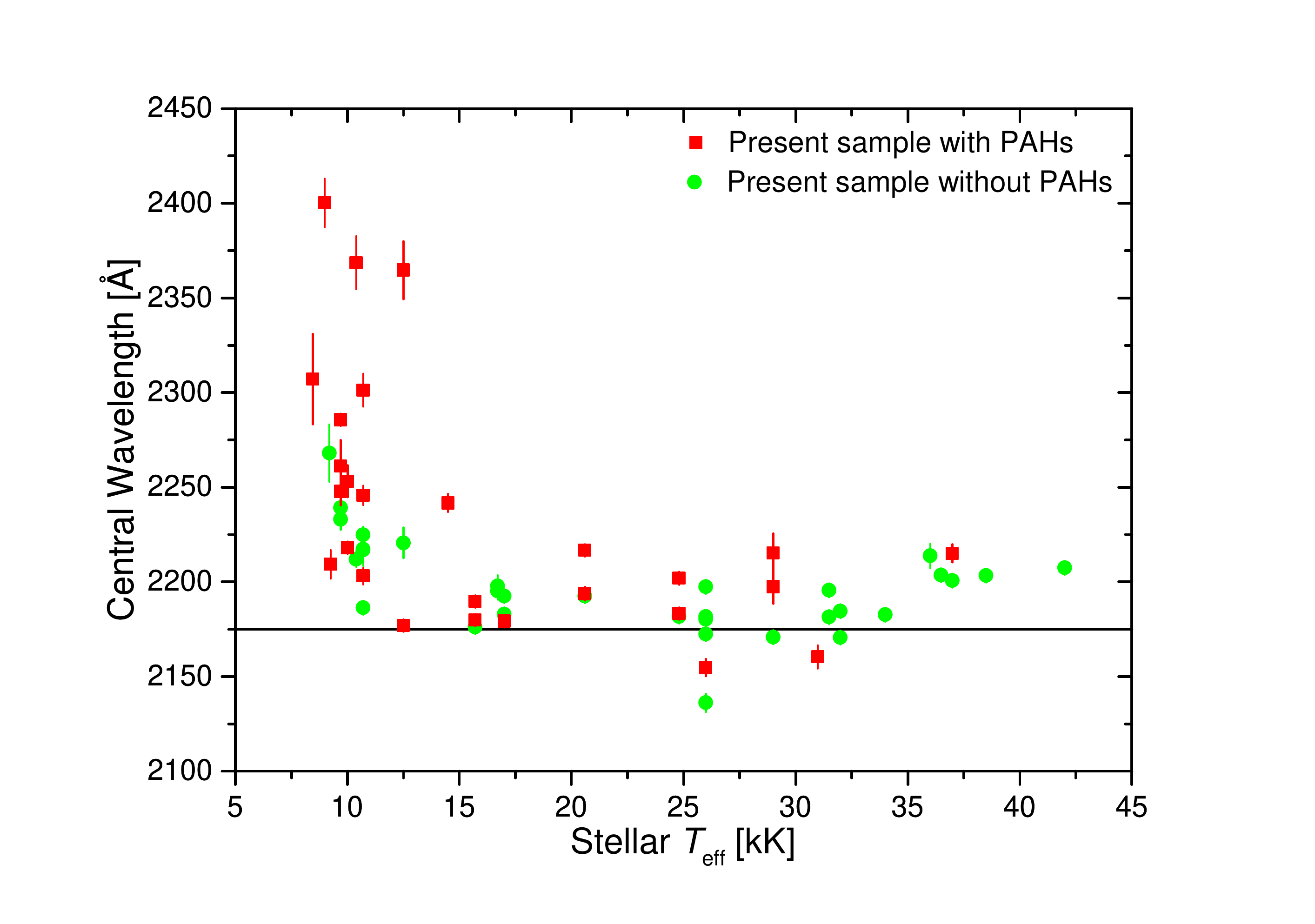}
	\caption{Directly measured central wavelength of the broad UV interstellar absorption feature along lines of sight to stars with different effective temperatures. All stars with $T_{\text{eff}}$ $>$ 14.5 kK lie within 2195$\pm$ 40\AA, which is consistent with the nominal 2175\AA\ value (horizontal line). In the cooler stars with 8 kK $< T_{\text{eff}} < 14.5$\,kK, the central UV wavelength can shift up to 2400 \AA}
	\label{UVvsT}
\end{figure}

Figure \ref{UVvsT} shows the position of the UV 
absorption feature $\lambda_0$ as a function of \Teff .
The longest wavelengths are observed in the relatively cool stars with 8.25 kK $<$ $T_{\text{eff}}$ $<$ 14.5 kK.
In these objects, the central UV wavelength can shift up to 2400 \AA , while the hotter stars with $T_{\text{eff}} > 14.5$\,kK approximately follow a nominal value of $2195 \pm 40$\,\AA. 
Furthermore, the most extreme (long) UV wavelengths are observed in those stars that also have PAH emission bands in the MIR. 
A Kolmogorov-Smirnov (KS) test on the cool stars from the main and control samples gives a probability of only 0.037 that they are drawn from the same underlying sample.

\begin{figure}[!]
	\hspace{-0.6cm}
	\includegraphics[width=0.56\textwidth]{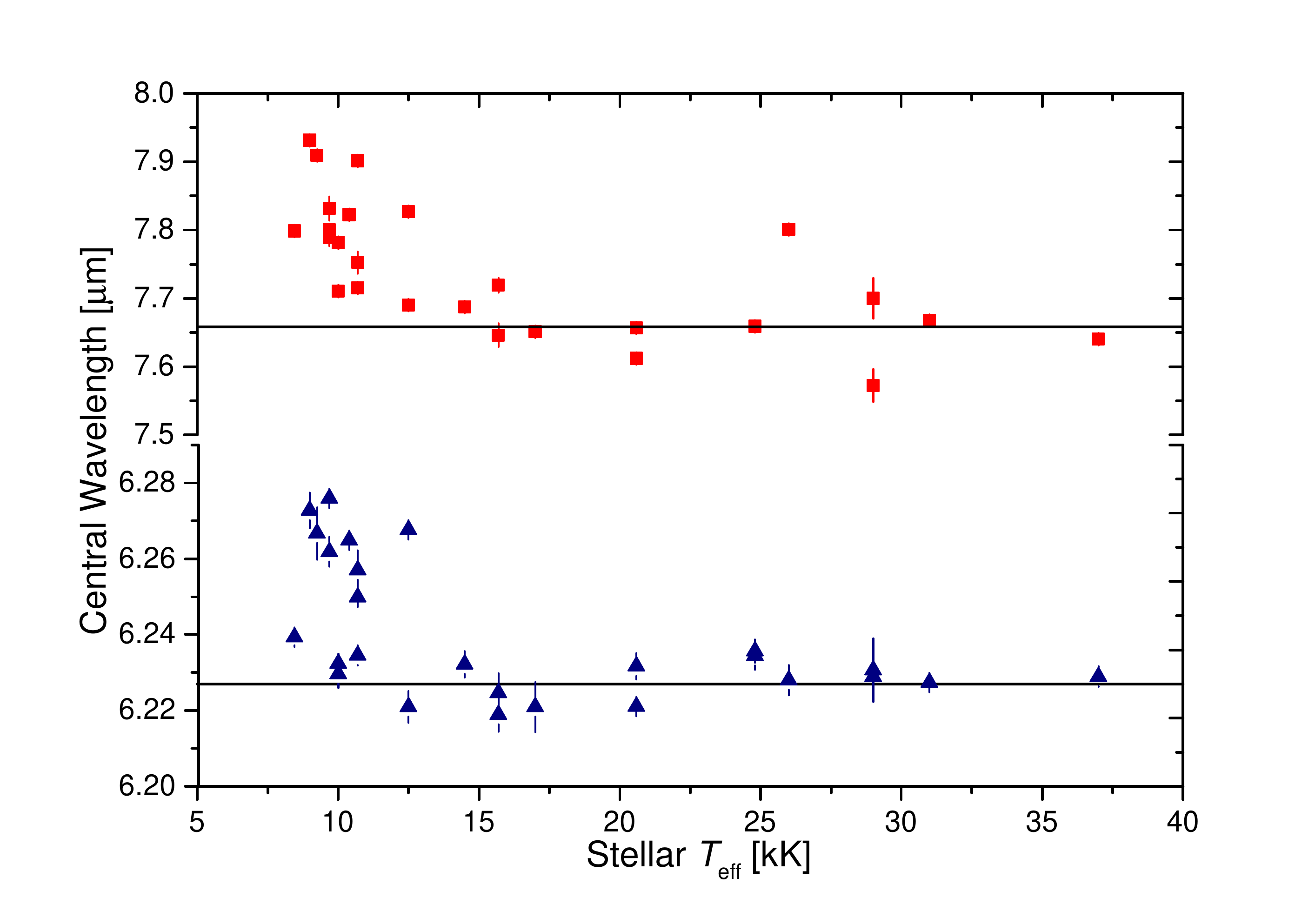}
	\caption{Central wavelength of the 7.7 $\mu$m and 6.2 $\mu$m PAH bands for stars from the present sample with different effective temperatures. Central wavelength of the PAH MIR bands in hot stars is consistent with the nominal positions at 6.23 $\pm$ 0.015 $\mu$m and at 7.66 $\pm$ 0.05 $\mu$m (horizontal lines). As the temperature drops below $T_{\text{eff}}  < 14.5$\,kK, the central wavelength gradually increases with decreasing temperature. }
	\label{MIRvsT}
\end{figure}

Figure \ref{MIRvsT} presents the central wavelengths of the MIR PAH emission bands. For the 6.2 $\mu$m and 7.7 $\mu$m emission bands, the longer MIR central wavelengths appear preferably in the cooler host stars, while hotter host stars feature nominal values of $6.23 \pm\ 0.015~\mu$m and $7.66 \pm\ 0.05~\mu$m. We find no significant shifts in the 8.6, 11.2, and 12.7 $\mu$m bands, in line with the results of \citet*{van Diedenhoven}.

\begin{figure}[!]
	\hspace{-1cm}
	\includegraphics[width=0.6\textwidth]{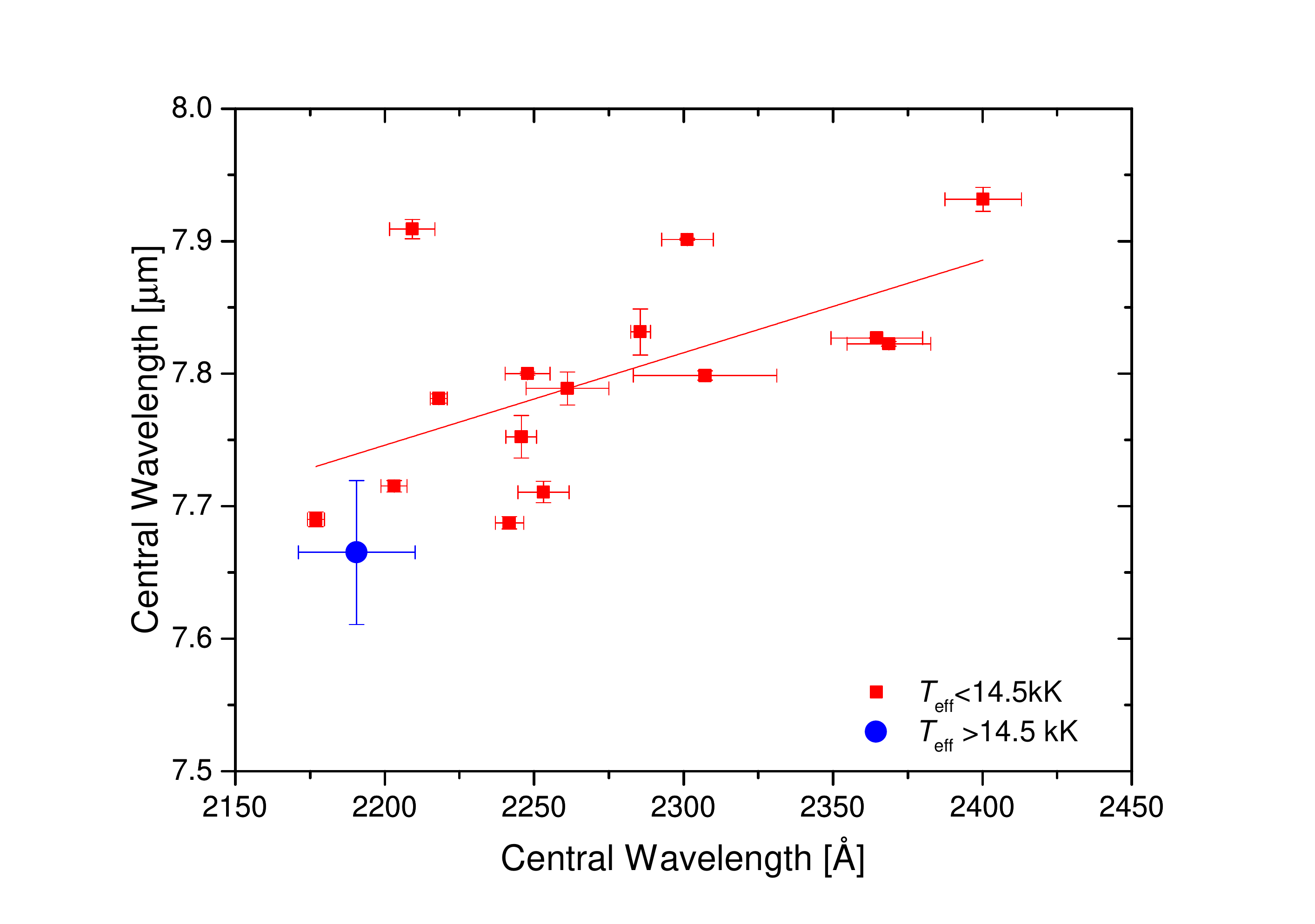}
	\caption{Anomalously long central wavelengths of the 2175 Å UV absorption feature and the 7.7$\mu$m PAH emission band. The lower temperature ($T_{\text{eff}} < 14.5$\,kK) part of the present sample (red squares) shows red-shifts of both UV and MIR features from their respective nominal regimes around 2195\,\AA\ and $7.66\,\mu$m (blue data point and error bars) where most stars, including all of the hotter ones, reside. }
	\label{UVIRcor}.
\end{figure}

Figure \ref{UVIRcor} presents the direct correspondence between the position of the UV absorption feature and that of the 7.7\,$\mu$m  emission band for stars with \Teff\ $< 14.5$\,kK. 
A reasonably good correlation is found between the two shifts, with Pearson's $R = 0.57$ and a chance probability of 0.021.
The correlation improves to $R = 0.77$ and a chance probability of $8.1 \times 10^{-4}$ if the significantly outlying star on the upper left side (HD\,141569) is ignored. 
The $\rm{EW}$ of HD\,141569 (170 \AA) is the second lowest in our sample 

\section{Discussion}
\label{discussion}

\subsection{UV shifts}
Previous UV samples \citep{FM86,FM88,FM90,FM05,FM07,CCM89,valancic2004} showed that the central wavelength  $\lambda_0$ was almost constant, with $\lambda_0 = 2178.5 \pm\ 9.1$\,\AA\ \citep{FM07}, and with a maximum range of 2130\,\AA\ to 2222\,\AA\ \citep{valancic2004}. The present sample, however, has a much broader and also shifted distribution of  $\lambda_0$. 
This is particularly the case for the stars that have PAH MIR emission bands, and less so for those that do not. 
Stars with and without MIR PAH emission, respectively, have a weighted mean of 2192.3\,\AA\ and 2189.5~\AA . 
The standard deviation of 63.9\,\AA\ is large for stars with MIR PAH emission, and less (24.2\,\AA ) for those without it.
The present sample has more shifted UV features than previous samples because it was selected differently.
We did not impose any visible extinction requirements, nor assume a standard extinction curve.
In fact, our sample includes many stars with low color excess that fitted locally show a strong UV feature.
In addition, our method allows the stellar metallicity to deviate from solar and thus improve the fit to the observed spectrum.

Few early claims in the literature of shifts towards long wavelength of the UV feature attributed it to carbon rich environments around the stars \citep{Buss  1,Greenstein  1,Hecht  1}.
We analyzed the spectra of all of these sources.
The objects HD\,89353 \& HD\,213985, discussed by \citet{Buss  1} are included in our sample, and both show strong PAH MIR emission (see Table~1). 
On the other hand, the Abell 30 planetary nebula \citep{Greenstein  1} and R\,Y Sgr \& R CrB \citep{Hecht  1}  do not have significant UV absorption, and as such do not qualify for inclusion in the present sample, nor can they represent regions of the ISM where UV absorption can be studied. 

\subsection{MIR shifts}
\begin{figure}[!]
	\hspace{-1cm}
	\includegraphics[width=0.6\textwidth]{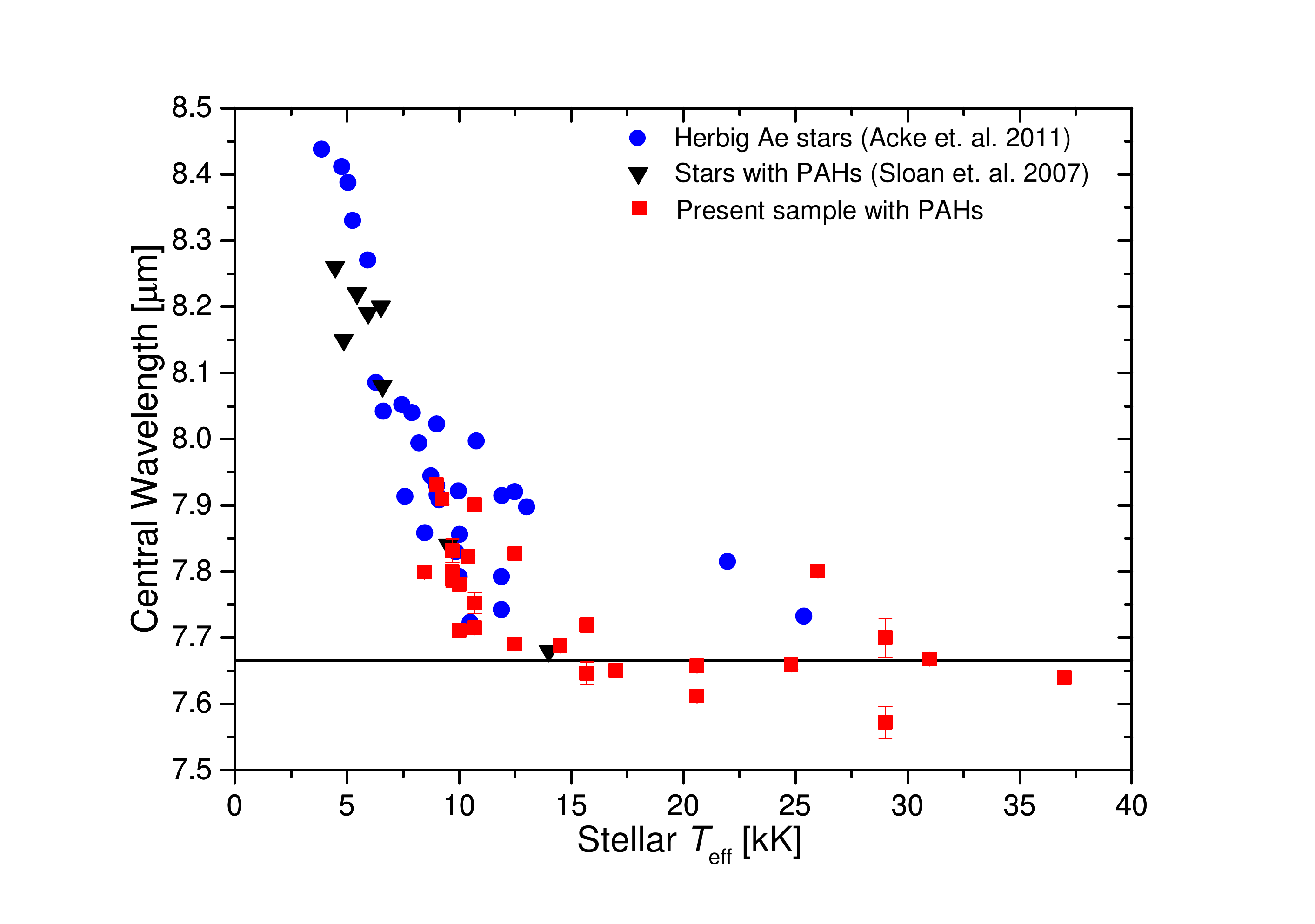}
	\caption{Central wavelength of the 7.7 $\mu$m MIR PAH band for stars with different effective temperatures. The dependence of the central wavelength on stellar temperature from the literature \citep{Acke 1,Sloan 1} is extended to higher temperatures of up to 37 kK using the present UV selected sample of hot stars. }
	\label{LMIRvsT}
\end{figure}

The dependence of the PAH MIR shifts on \Teff\ in Fig.~\ref{MIRvsT} corroborates previously published results, as can be seen in Figure \ref{LMIRvsT}, where the position of the 7.7\,$\mu$m band as a function of $T_{\text{eff}}$ continues to increase down to 4\,kK \citep[data from][]{Acke 1,Sloan 1}. 
The present sample extends the trend to higher temperatures and demonstrates that he central wavelength tends to the nominal value of $\sim 7.66\,\mu$m above 15\,kK.
This result is not different than the findings of \citet{Peeters} that the MIR PAH central wavelengths change with stellar type (and hence temperature).
Note that stellar temperature can not be the only parameter governing the MIR shift, as the 7.7 $\mu$m feature around planetary nebulae (PNe), where \Teff\ can reach 100 kK, is also shifted to 7.9 $\mu$m \citep{joblin 2}.

The limitation of the statistical correlation notwithstanding, the important result here is that stars whose UV absorption is red-shifted also have red-shifted MIR PAH bands, both beyond the high-\Teff\ average (blue data point in Fig.~\ref{UVIRcor}), suggesting that PAHs are responsible for at least the shifted UV absorption. 

The correlation presented in  Figure \ref{UVIRcor} shows that many IR-shifted sources in the sample appear specifically at $7.9\,\mu$m. This feature was identified in proto-planetary disks \citep{Boersma 1} and UV-bright PNe with large PAHs, while here it is found in the weakest UV (low \Teff ) sources. The scatter in the correlation of Figure \ref{UVIRcor} is likely due to more complicated molecular astro-chemistry and the varying circumstellar environments, but the upshot is that stars that feature red-shifted UV absorption also feature red-shifted MIR emission bands. 
This result, in conjunction with the same pattern of the UV absorption and the MIR emission shifting with decreasing stellar temperature, suggests that the UV absorption feature around 2200\,\AA\ is a result of the same molecules that emit the MIR bands, and which are believed to be PAHs. 

\subsection{What drives the UV and MIR wavelength shifts?} 
Longer UV and MIR wavelengths indicate larger PAH molecules with more than 50 C-atoms \citep{Steglich 1,Steglich 2,Cami 1,Ricca 1,Bauschlicher 1}, as can also be seen by comparing, e.g., cross sections of Coronene C$_{24}$H$_{12}$ vs. Circumcoronene C$_{54}$H$_{18}$, or Ovalene C$_{32}$H$_{14}$ vs. Circumovalene C$_{66}$H$_{20}$ \citep{Malloci 1}. 
Red-shifted MIR bands tend to arise from larger molecules as well, but ionization \citep{Malloci 1}, nitrogen atoms in the aromatic rings \citep*{Hudgins 1}, and aliphatic compounds \citep{Sloan 1} can also shift their central wavelength.

Although PAH UV shifts have not been studied systematically, and although several effects can cause MIR shifts, the present correlation between the two, and the similar dependence on $T_{\text{eff}}$ lends support to the size playing a dominant role. If this interpretation is correct, larger molecules are present around the cooler stars, but not around the hot ones, with $T_{\text{eff}} \sim15$\,kK being the dividing temperature. 

\subsection{What is special about 15\,kK?}
Both the UV and the MIR features start being significantly shifted only below $\sim 15$\.kK.
This is the critical temperature for UV-pressure driven stellar winds \citep*{Prinja  1,Lamers 1,Kudritzki  1}. 
Although one might expect the UV flux from the hot stars to destroy the smaller molecules, this is not what is observed here.
Another possible effect of the hot-star winds may have to do with the shocks and the hot (MK) X-ray gas that they produce.
If shock heating destroys preferentially the large molecules, it could explain why no UV and MIR wavelength shifts are observed in these hot stars, where the nominal wavelengths of 2175\,\AA\ and $7.66\,\mu$m may represent normal ISM composition of dust grains and normal-size PAHs with $\lesssim$50 C-atoms \citep{Tielens1,Sellgren 1}.
This speculation remains to be investigated in future studies.


\subsection{Is the phenomenon circumstellar or interstellar?}
One may wonder if the absorbing and emitting molecules are associated with the background stars, or are they typical of mean ISM properties.
The similar behavior  of the  UV and IR wavelength shifts, as well as the dependence on stellar temperature strongly suggest these features are associated with the host stars. 
Moreover, we find no correlation between the measured UV $\rm{EW}$ and the Galactic latitude (a correlation coefficient of determination R$^2$ = 0.14), or the EW and the stellar distance (R$^2$ = 0.09). This is in fact expected for the present sample that consists mostly of nearby stars, and for which the measured UV and MIR properties are not associated with mean ISM values along the random lines of sight. 

One can estimate mean ISM UV absorption and IR emission based on the expected absorbed and emitted fluxes in the ISM \citep{Siebenmorgen}. The mean ISM H number density $n_{\text H}$ along a random line of sight can be expected to range between 0.1 to 1.0 cm$^{-3}$. Assuming a solar C abundance of C/H\,$= 2.4 \times 10^{-4}$, and that 10\% of the C is in PAHs, the number density of C atoms in PAHs is $n_{\text PAH_C} = 2.4 \times 10^{-5}n_{\text H}$.
The column density towards a star is then the mean density times the distance to the star.
The most distant stars in our sample are $\sim$ 2 kpc ($6.2 \times 10^{21}$ cm) away, which corresponds to column densities of $N_{\text H} = 6.2 \times 10^{21}n_{\text H}$\,cm$^{-2}$ and  $n_{\text PAH_C} = 1.5 \times 10^{17}n_{\text H}$.

The UV PAH absorption cross-section at the center of the feature ($\lambda_0$) is $\sigma$ = 10$^{-17}$ cm$^{2}$ per PAH C atom \citep{Siebenmorgen}. Hence, the mean ISM optical depth is expected to be $\tau (\lambda_0) = \sigma n_{\text PAH_C} d$, where $d$ is the distance to the star. If $d_{\text pc}$ is the distance measured in pc, one can plug in the numbers and write simply  

\begin{equation}
\tau (\lambda_0) = 7.4 \times 10^{-4} n_{\text PAH_C} d_{\text pc}
\end{equation}

\begin{figure}[!]
	\hspace{-1cm}
	\includegraphics[width=0.6\textwidth]{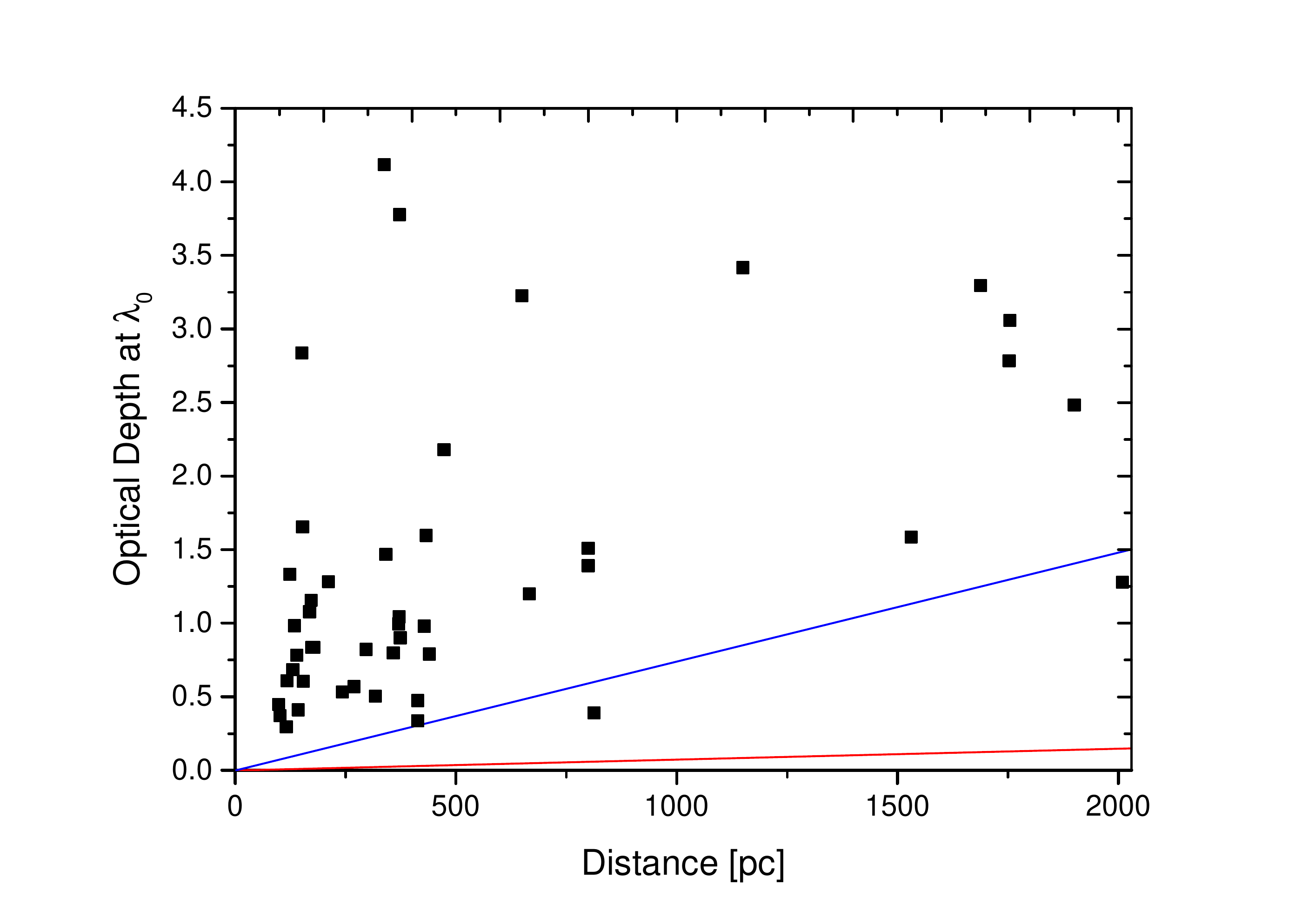}
	\caption{UV optical depth as a function of distance. The optical depth at the center of the UV feature is plotted as a function of the distance for 49 out of the 59 stars in the present sample. Three stars fall outside the plot (one with $\tau > 7$ and two with $d > 2.5$\,kpc), and for seven we could not find the distance. An analytical (linear) estimate for $\tau$ is plotted for $n_{\text H} = 0.1$\,cm$^{-3}$ (lower line) and for 1.0 cm$^{-3}$ (upper line). The mean ISM optical depth is expected to fall between these two lines, demonstrating that the UV absorption measured in the sample is not commensurate with ISM absorption and can be mostly ascribed for the most part to the stellar environment.}
	\vspace{0.2cm}
	\label{OD}
\end{figure}

\noindent We were able to find distances to 51 out of the 59 stars in the sample, most of which are closer than 500\,pc and therefore are not expected to have much ISM absorption. In Figure \ref{OD}, we compare the analytic estimate with the optical depth $\tau (\lambda_0) = A/ (\pi w^2)$ measured from the fitted spectra. It is demonstrated that all of the current measurements lie much above the line for $n_{\text H} = 0.1$\,cm$^{-3}$, and all but two stars lie above the line for $n_{\text H} = 1$\,cm$^{-3}$.

In the IR, the intensity of the $7.7\,\mu$m band per H column density is $\nu I_{\nu}/N_{\text H} = 5.6 \times 10^{-26}$\,erg s$^{-1}$ cm$^{-2}$str$^{-1}$ \citep*{Siebenmorgen}.
Using N$_H$ from above, the intensity up to 2\,kpc would be $\nu I_\nu = 3.1 \times 10^{-4}n_{\text H}$\,erg\,s$^{-1}$\,cm$^{-2}$str$^{-1}$. 
The \spitzer /IRS spectra we use are reduced from a 20 arcsec$^{2}$ extraction (4.7 x 10$^{-10}$ str) on the sky. 
Thus, the expected mean ISM flux in the $7.7\,\mu$m band is 

\begin{equation}
\nu F_\nu = 1.5 \times 10^{-13}n_{\text H}~\text{erg\,s}^{-1}\text{cm}^{-2} 
\end{equation}

\noindent The integrated flux of the feature is approximately $F_\nu \Delta \nu= \nu F_\nu (\Delta \nu/ \nu) =\nu F_\nu(\Delta \lambda/ \lambda) $.
The measured width (FWHM) of the $7.7\,\mu$m feature is on average $\Delta \lambda = 0.6\,\mu$m, or $\Delta \lambda / \lambda = 0.078$, thus the expected flux of the feature due to the ISM emission is $F_\nu \Delta \nu = 1.2 \times 10^{-14}n_{\text H}$\,erg\,s$^{-1}$cm$^{-2}$, even for a distant target 2\,kpc away. 
For the range of n$_{\text H}$ between 0.1 - 1.0 cm$^{-3}$, we thus obtain $F_\nu \Delta \nu \approx 10^{-15} - 10^{-14}$\,erg\,s$^{-1}$\,cm$^{-2}$, which is by far weaker than the flux range of $3.5 \times 10^{-13}$ to $2 \times 10^{-9}$\,erg\,s$^{-1}$\,cm$^{-2}$ measured in our sample.
We conclude therefore that the UV and IR measurements towards stars in the present sample can mostly be associated with the stellar environment and not with the random, mean ISM properties of the individual lines of sight.

\section{Summary \& Conclusions}
\label{Conclusions}

Our results can be summarized as follows:

\begin{itemize}
\item {We find significant shifts of the UV extinction feature around 2200\,\AA .}
\item {These shifts become more appreciable as the temperature of the background star decreases below 15\,kK where hot-star winds cease to exist.}
\item {Exactly the same trend of shifts with stellar temperature occurs also in MIR PAH bands.}
\item{We find the shifts in the UV to correlate with those in the MIR, suggesting they both originate in the physical properties of PAH molecules.}
\item{One possible reason for the shifts could be molecular size.}
\end{itemize}

These results open new possibilities to diagnose specific molecular species in the ISM, which is unattainable in the MIR alone, but could be achieved with more UV laboratory and computational experiments. These could test whether the shifts are due to molecular size. While the present results associate large PAHs preferably with windless, planet-forming stars, the astrophysical setting for the formation and survival of these organic molecules, and their possible connection to the origin of life remains to be investigated. Moreover, our findings  will allow one to predict the presence of PAHs in the ISM and the positions of their MIR bands from the UV spectra, and vice versa. Furthermore, the observed variation in the 2175\AA\ may be the long sought-after electronic signature of PAHs in the UV.

\acknowledgments Work at the Technion was supported by the I-CORE program of the Planning and Budgeting Committee (grants 1937/12 \& 1829/12). 
A.B. acknowledges programming support from Uria Peretz.
E.B received funding from the European Unions Horizon 2020 research and innovation programme under the Marie Sklodowska-Curie grant agreement no. 655324.

\bibliographystyle{apj}

\begin{thebibliography}
 
\bibitem[Allamandola,Tielens \& Barker(1989)]{Allamandola89}
Allamandola, L. J., Tielens, A. G. G. M., Barker, J. R., 1989, ApJS, 71, 733A
\bibitem[Acke et al.(2010)]{Acke 1}
Acke, et~al., 2010, ApJ, 718, 558
\bibitem[Bauschlicher et~al.(2010)]{Bauschlicher 1}
Bauschlicher, C. W, et~al., 2010, ApJS, 189, 341
\bibitem[Berné et~al. (2009)]{Berné 1}
Berné, C.,et~ al., 2009, A\&A, 495, 827
\bibitem[Boersma, et al.(2009)]{Boersma 1}	
Boersma, C.,at .al.,2009,A\&A,502,175
\bibitem[Buss et~ al.(1989)]{Buss  1}	
Buss, Richard H., Jr., Snow, Theodore P., Jr., Lamers, Henny J. G. L. M.,1989,ApJ,347,977
\bibitem[Cami(2011)]{Cami 1}
Cami, J., 2011, EAS, 46, 117
\bibitem[Castelli \& Kurucz (2004)]{Castelli 1}
Castelli, F. \& Kurucz, R.L., 2004, astro-ph/ 0405087
\bibitem[Cardelli et~ al.(1989)]{CCM89}
Cardelli, J. A., Clayton, G. C. \&Mathis, J. S., 1989, ApJ,  345, 245
\bibitem[deJager \& Nieuwenhuijzen (1987)]{dejager}
deJager, C., Nieuwenhuijzen, H.,1987, A\&A, 177, 217
\bibitem[Draine(2003)]{Draine  1}
Draine, B. T., 2003, ARA\&A, 41, 241
\bibitem[Ehrenfreund, et~ al.(2006)]{Ehrenfreund 1}
Ehrenfreund, P.,et~al. 2006, AsBio, 6, 490
\bibitem[Fitzpatrick \& Massa(1986)]{FM86}
Fitzpatrick, E. L.,\& Massa, D., 1986, ApJ, 307, 286
\bibitem[Fitzpatrick \& Massa(1988)]{FM88}
Fitzpatrick, E. L.,\& Massa, D., 1988,  ApJ, 328, 734
\bibitem[Fitzpatrick \& Massa(1990)]{FM90}
Fitzpatrick, E. L.,\& Massa, D., 1990,  ApJ, 72, 163
\bibitem[Fitzpatrick \& Massa(2005)]{FM05}
Fitzpatrick, E. L.,\& Massa, D., 2005,  ApJ, 130, 1127
\bibitem[Fitzpatrick \& Massa(2007)]{FM07}
Fitzpatrick, E. L.,\& Massa, D., 2007,  ApJ, 663, 320
\bibitem[Galliano et~ al.(2008)]{Galliano}
Galliano, et~ al.,  2008,ApJ,679,310
\bibitem[Greenstein(1981)]{Greenstein  1}
Greenstein, J. L.,1981,ApJ,245,124
\bibitem[Hecht et~ al.(1984)]{Hecht  1}
Hecht, J. H.; Holm, A. V.; Donn, B.; Wu, C.-C.,1984,ApJ,280,228
\bibitem[Hony, et~ al.(2001)]{Hony 2006}
Hony, et~ al. 2001, A\&A,370,1030
\bibitem[Hudgins et~ al. (2005)]{Hudgins 1}
Hudgins, D.M., Bauschlicher, C.W., Allamandola, L. J., 2005, ApJ 632, 316
\bibitem[Joblin, et~al.(1992)]{joblin 1}
Joblin, C., Léger, A., Martin, P.,1992, ApJ, 393, L79
\bibitem[Joblin et~al.(2008)]{joblin 2}
Joblin, C.,et~ al., 2008, A\&A, 490, 189
\bibitem[Kudritzki et~ al.(2000)]{Kudritzki  1}	
Kudritzki, Rolf-Peter, Puls, Joachim,2000,ARA\&A,3,613
\bibitem[Lamers et~ al.(1995)]{Lamers 1}
Lamers, H.J.G.L.M., Snow,T.P., Lindholm,D.M.,1995, ApJ 455, 269 
\bibitem[Léger \& Puget (1984)]{leger 1}
Léger, A. \& Puget, J., 1984, Astron. Astrophys. 137, L5-L8
\bibitem[Li \& Draine(2002)]{lidr2002}
Li, A., Draine, B. T.  2002, ApJ, 572, 232
\bibitem[Malloci et~ al.(2007)]{Malloci 1}
Malloci, G., Joblin, C. , Mulas, G., 2007, Chem. Phys. 332, 353
\bibitem[Mori et al (2012)]{mori1}
Mori, Tamami I. et al,	2012,ApJ,744,68
\bibitem[Pecaut \& Mamajek(2013)]{Pecaut}
Pecaut, Mamajek, 2013,ApJS,208,9
\bibitem[Pino et. al.(2008)]{pino}
Pino, et al.,2008,A\&A,490,665
\bibitem[Peeters et. al.(2002)]{Peeters}
Peeters, E., et al.,2002, A\&A,390,1089
\bibitem[Puget \& Léger (1989)]{Puget 1}
Puget, J. L., Léger, A., ARA\&A,1989, 27, 161,
\bibitem[Prinja et~al.(1998)]{Prinja  1}	
Prinja, Raman K., Massa, Derck L.,1998,ASPC,131,218
\bibitem[Ricca et~al.(2012)]{Ricca 1}
Ricca, A.,et~al., 2012, ApJ, 754, 75
\bibitem[Savage et al.(1985)]{Savage85}
Savage et al, 1985, ApJ, 59, 397
\bibitem[Schmidt-Kaler (1982)]{Schmidt-Kaler}
Schmidt-Kaler,1982, Http://obswww.unige.ch/gcpd/mk01bv.html
\bibitem[Siebenmorgen et~al.(2014)]{Siebenmorgen}
Siebenmorgen, R., Voshchinnikov, N.V., Bagnulo, S., 2014, A\&A, 561A, 82
\bibitem[Sellgren (1984)]{Sellgren 1}
Sellgren, K., 1984, ApJ, 277, 623
\bibitem[Sloan et~al.(2003)]{Sloan 2}
Sloan et al., 2003, ApJS, 147, 379
\bibitem[Sloan et al.(2007)]{Sloan 1}
Sloan,  G. C, et al., 2007, ApJ, 664, 1144
\bibitem[Smith, Clayton \& Valencic (2004)]{smith2004}
Smith, T.L.,Clayton, G.C.,Valencic, L.,2004,AJ,128,357
\bibitem[Stecher(1965)]{Stecher 1}
Stecher,T.P., 1965, ApJ, 142, 1683
\bibitem[Stecher \& Donn(1965)]{Stecher 2}
Stecher,T.P.\& Donn, B. , 1965, ApJ, 142, 1681
\bibitem[Steglich et~al.(2010)]{Steglich 1}
Steglich, M. ,et~al.,2010, ApJ, 712, 16
\bibitem[Steglich et~al.(2011)]{Steglich 2}
Steglich, M.,et~al., 2011, ApJ, 742, 2
\bibitem[Tielens(2008)]{Tielens1}
Tielens, A. G. G. M.,2008,ARA\&A,46,289
\bibitem[Uchida et al.(2000)]{Uchida}
Uchida, et al.,2000,ApJ,530,817
\bibitem[Valencic et~al.(2004)]{valancic2004}
Valencic, L. A., Clayton, G. C. \& Gordon, K. D.,2004,ApJ,616,912
\bibitem[Van Diedenhoven et~al.(2004)]{van Diedenhoven}
Van Diedenhoven, B, et~al ,2004, ApJ, 611, 928
\bibitem[Zubko et~al.(2004)]{zubko  1}
Zubko, V., Dwek, E., Arendt, R.G.,2004, ApJS, 152, 211
\end{thebibliography}

\end{document}